\documentclass[preprint,12pt]{elsarticle}

\usepackage{amssymb}
\usepackage{amsmath}

\usepackage{algorithm}
\usepackage{algpseudocode}
\usepackage{url}

\usepackage{float}

\journal{Computational Physics}

\begin{document}

\begin{frontmatter}


\title{Reduced Floating-Point Precision Implicit Monte Carlo\tnoteref{LA-UR}}
\tnotetext[LA-UR]{LA-UR-25-30524}
\author[OSU]{Simon Butson\corref{cor1}} 
\cortext[cor1]{butsons@oregonstate.edu}
\author[LANL]{Mathew Cleveland}
\author[LANL]{Alex Long}
\author[OSU]{Todd Palmer}

\affiliation[OSU]{organization={Oregon State University},
             city={Corvallis},
             state={Oregon},
             country={}}


\author{} 

\affiliation[LANL]{organization={Los Alamos National Lab},
            city={Los Alamos},
            state={New Mexico},
            country={}}

\begin{abstract}
This work demonstrates algorithms to accurately compute solutions to thermal radiation transport problems using a reduced floating-point precision implementation of the Implicit Monte Carlo method. Several techniques falling into the categories of arithmetic manipulations and scaling methods are evaluated for their ability to improve the accuracy of reduced-precision computations. The results for half- and double-precision implementations of various thermal radiation benchmark problems are compared.
\end{abstract}




\end{frontmatter}



\section{Introduction}
\label{sec1}

Thermal radiative transfer poses a challenging and interesting application area for reduced-precision calculations. The nonlinear physics with physical quantities spanning many orders of magnitude poses immense difficulties for accurate reduced-precision calculations. In this work, we consider a reduced-precision implementation of the Implicit Monte Carlo method (IMC) for solving radiative transfer problems \cite{FLECK1971313}.  To successfully implement the IMC method in reduced-precision and produce accurate results requires the use of two broad categories of techniques, arithmetic manipulations, and scaling methods. Arithmetic manipulations include things such as using alternate numerical summation algorithms with an algorithmic procedure that changes the order of operations when calculating the product of multiple values. Scaling techniques can be static or dynamic and include ideas such as changing base units, rescaling distance on a relative basis, and using multiple algorithmically selected energy scales. The overarching goal of all these techniques is to keep reduced-precision variables in a suitable numerical range to prevent the occurrence of underflow/overflow and to minimize the effects of round-off errors in calculations. The motivation for this work is two-fold; to find techniques to accelerate scientific calculations using reduced-precision floating-point data-types and to safeguard against loss of simulation capabilities if future computing architectures lack support for higher-precision floating-point data types. This work builds on earlier investigations at Los Alamos National Laboratory on the implementation of the IMC method in reduced floating-point precision~\cite{alexlong24}. In particular, we extend the previous distance scaling method to work in conjunction with multiple energy scales and replace the ternary product order of operations rearranger with a more robust method that can accept an arbitrary number of values and scale factors. 
The methods and results of this work were implemented using the Julia programming language~\cite{bezanson2017julia}; the developed software is accessible at the following link: \url{https://github.com/simonbutson/MixedPrecisionIMC.jl}.

\subsection{Thermal Radiation Transport \& The Implicit Monte Carlo Method}
\label{subsec1}
The transport of thermal radiation in a system depends on the exchange of energy between absorbing/emitting materials and photons in a radiation field. Absorbed photons add energy to materials and emitted photons decrease material energy. The physical processes involved in thermal radiative transfer can be modeled mathematically using coupled radiative transfer and material energy balance equations. The radiation transport is described by the Boltzmann transport equation which is written for the photon specific intensity as a function of space $\textbf{r}$, angle $\Omega$, time $t$, and frequency $\nu$:
\begin{equation}
    \begin{array}{c}
    \frac{1}{c}\frac{\partial I(\boldsymbol{r},\boldsymbol{\Omega}, \nu, t)}{\partial t} + \boldsymbol{\Omega} \cdot \nabla I(\boldsymbol{r}, \boldsymbol{\Omega}, \nu, t) + \sigma_a (\boldsymbol{r},\nu,T) I(\boldsymbol{r},\boldsymbol{\Omega},\nu,t) \\
    \\
    =\sigma_a(\boldsymbol{r},\nu,T) B(\nu,T) +\frac{Q_r(\boldsymbol{r}, \nu,t)}{4 \pi}. 
    \end{array}
\end{equation}
This equation is coupled to the material energy balance:
\begin{equation}
    \begin{array}{c}
     \frac{\partial U_m(\boldsymbol{r},t,T)}{\partial t} = \int_0^\infty \int_{4\pi} \sigma_a(\boldsymbol{r},\nu,T)I(\boldsymbol{r},\boldsymbol{\Omega}, \nu, t)d\Omega d\nu  \\
     \\
     -\int_0^\infty \sigma_a(\boldsymbol{r},\nu,T) B(\nu,T) d\nu + Q_m(\boldsymbol{r},t),
     \end{array}
\end{equation}
where the material energy density $U_m(\textbf{r},t,T)$ is defined in relation to the heat capacity $c_v(\boldsymbol{r},T)$ as: 
\begin{equation}
    U_m(\boldsymbol{r},T) = \int_0^Tc_v(\boldsymbol{r},T')dT',
\end{equation}
and the equilibrium radiation energy density $U_r(\textbf{r}, T)$ is:
\begin{equation*}
    U_r(\boldsymbol{r},T) = aT_r(\boldsymbol{r},t)^4
\end{equation*}
where $a$ is the radiation constant:
\begin{equation}
    a= \frac{8\pi^5k_{B}^4}{15hc^3},
\end{equation}
and $T$ and $T_r$ are the material and radiation temperatures respectively. Additional inhomogeneous radiation and material energy sources $Q_r(\boldsymbol{r}, \nu, T)$ and $Q_m(\boldsymbol{r,t})$ may also be present.
These equations are also subject to initial conditions describing the starting state of the system, and boundary conditions which we can write generically as:
\begin{gather}
    I(\boldsymbol{r}, \boldsymbol{\Omega}, \nu, 0) = I_0(\boldsymbol{r},\boldsymbol{\Omega},\nu), \\
    T(\boldsymbol{r}, 0) = T_0(\boldsymbol{r}), \\
    I(\boldsymbol{r}, \boldsymbol{\Omega},\nu,t) = I_B(\boldsymbol{r}, \boldsymbol{\Omega}, \nu,t), \ \text{for} \ \boldsymbol{r} \in \partial V \ \text{and} \ \boldsymbol{\Omega}\cdot \boldsymbol{n} < 0
\end{gather}
An important assumption used in these equations is that the matter is in ``local thermodynamic equilibrium" (LTE) and that its emissivity can be described by a black body with temperature $T$. Additional assumptions made are to neglect complicating physics such as different ion-photon interactions, hydrodynamic motion,
thermal conduction, and relativistic effects~\cite{40IMC}. The Planck spectrum for a black body is given as:
\begin{equation}
    B(\nu, T) = \frac{2h\nu^3}{c^2}\frac{1}{\exp(\frac{h \nu}{k_BT})-1}.
\end{equation}
Commonly used units for radiative transfer applications and high energy density physics are Jerks (Jk) = $1\times10^9$ J for energy and shakes = $1\times 10^{-8}$ s for timing. A list of parameters used in thermal radiative transfer calculations is shown in Table \ref{tab:IMCparams}. 

\begin{table}[H]
    \centering
    \resizebox{\textwidth}{!}{
    \begin{tabular}{|c|c|c|}
       \hline
        \textbf{Symbol} &  \textbf{Parameter} & \textbf{Value/Units} \\
        \hline
         $I = ch\nu n(\boldsymbol{r}, \boldsymbol{\Omega}, \nu,t)$ & Specific Intensity & Jk/$\text{cm}^2$-ns-keV-sr \\ 
         \hline
         $c$ & Speed of Light & $29.9792$ cm/ns \\ 
        \hline
        $h$ & Planck's Constant & $4.13567 \times 10^{-9}$ keV-ns \\ 
        \hline
        $\nu$ & Frequency & 1/ns \\
        \hline
        $n(\boldsymbol{r}, \boldsymbol{\Omega}, \nu, t)$ & Mean Photon Population Per Unit of Phase Space & photons/$\text{cm}^3$-sr \\
        \hline
        $c_v(\boldsymbol{r},T)$ & Heat Capacity & Jk/cc-ns-keV-sr \\
        \hline
        $k_B$ & Boltzmann Constant & $8.617 \times 10^{-8}$ keV/K \\
        \hline
        $T(\boldsymbol{r},t)$ & Material Temperature & keV \\
        \hline
        $T_r(\boldsymbol{r},t)$ & Radiation Temperature & keV \\
        \hline
        $B(\nu,T)$ & Planck spectrum & Jk/$\text{cm}^2$-ns-keV-sr \\
        \hline
        $\sigma_a(\boldsymbol{r}, \nu, T)$ & Absorption Opacity & $\text{cm}^{-1}$ \\
        \hline
        $a$ & Radiation Constant & 0.01372 Jk/$\text{cm}^3\text{-keV}^4$ \\
        \hline
        $Q_r(\boldsymbol{r}, \nu, t)$ & Radiation Source & Jk/$\text{cm}^3$-ns-keV \\
        \hline
        $Q_m(\boldsymbol{r}, t)$ &  Material Source & Jk/$\text{cm}^3$-ns \\
        \hline
    \end{tabular}
    }
    \caption{IMC Parameters Table}
    \label{tab:IMCparams}
\end{table}

A standard solution method for thermal radiative transfer problems is the Implicit Monte Carlo method (IMC) introduced in 1971 by Fleck \& Cummings \cite{FLECK1971313}. In IMC, the the thermal radiation transport equation is linearized  to yield a set of discrete radiation transport and radiation-material coupling equations. These equations can be numerically solved using a Monte Carlo simulation of particles (radiation energy bundles) that are emitted and absorbed by materials in the problem at discrete time-steps $t_n$. The IMC method introduces a Fleck factor $f$, corresponding physically to the absorption and isotropic re-emission of radiation within a time-step. The Fleck factor allows for effective scattering to occur, permitting larger time-steps to be used without instabilities arising in the Monte Carlo radiative transfer simulation. The IMC approach makes use of the ratio between the material and radiation energy density which is defined as:
\begin{equation}
    \beta(\boldsymbol{r},t) = \frac{\partial U_r}{\partial U_m} = \frac{dU_r/dT}{dU_m/dT} = \frac{4aT^3}{c_v(\boldsymbol{r},T)}
\end{equation}
Working through the full derivation in \cite{FLECK1971313} yields the expression for the Fleck factor:
\begin{equation}
    f = \frac{1}{1+\alpha c \sigma_{a,p}^n \Delta t \beta^n}
\end{equation}
where $\alpha \in [0,1]$ is an implicitness parameter of the time-step differencing and $\sigma_{a,p}$ is the Planck weighted opacity:
\begin{equation}
    \sigma_{a,p}=\frac{\int_0^\infty \sigma_a(\nu,T)B(\nu,T)d\nu}{\int_0^\infty B(\nu,T)d\nu} = \frac{\int_0^\infty \sigma_a(\nu,T)B(\nu,T)d\nu}{cU_r(T)}
\end{equation}
The IMC version of the thermal radiation transport equation is:
\begin{equation}
\begin{array}{c}\label{eq:radtransport}
        \frac{1}{c}\frac{\partial I(\boldsymbol{r},\boldsymbol{\Omega}, \nu, t)}{\partial t} + \boldsymbol{\Omega} \cdot \nabla I(\boldsymbol{r}, \boldsymbol{\Omega}, \nu, t) + \sigma_a^n I(\boldsymbol{r},\nu,T,t) = 
        \chi(\nu)\sigma_{a,p}^n fU_r^n \\
        \\
        + \chi(\nu)(1-f)\Big(\int\limits_0^\infty \int\limits_{4\pi} \sigma_a^n(\boldsymbol{r},\nu,T) I(\boldsymbol{r},\boldsymbol{\Omega},\nu,t)d\Omega'd\nu + Q_r(\boldsymbol{r}) \Big) 
        \end{array}
\end{equation}
and the IMC material energy balance equation is:
\begin{equation}
\begin{array}{c}\label{eq:matengbalance}
    \frac{U^{n+1}_m(\boldsymbol{r},T)-U_m^n(\boldsymbol{r},T)}{\Delta t} 
    = \frac{1}{\Delta t} \int\limits_{t_n}^{t_{n+1}} \int\limits_0^\infty \int\limits_{4\pi} f \sigma_a^n(\boldsymbol{r},\nu,T)I(\boldsymbol{r},\boldsymbol{\Omega},\nu,t) d\Omega'd\nu dt \\
    \\
    - f c \sigma_{a,p} U_r^n(\boldsymbol{r},T) + f Q_m(\boldsymbol{r},t)
    \end{array}
\end{equation}
where $\chi$ is the emission spectrum:
\begin{equation}
    \chi(\nu,T) = 
    \frac{\sigma_a(\nu,T)B(\nu,T)}{\int\limits_0^\infty \sigma_a(\nu,T) B(\nu,T) d\nu}
\end{equation}
While the above discussion gives the IMC equations in full generality, the results of this paper will make use of the gray approximation that replaces any frequency-dependent opacities with the Planck/gray opacity $\sigma_{a,p}$. This approximation avoids many of the numerical difficulties that arise when reduced-precision floating point numbers are used to represent frequency-dependent opacities, while still allowing for a variety of thermal radiation physics to be faithfully simulated.

\subsection{Implicit Monte Carlo Algorithm}
An overview of the Implicit Monte Carlo algorithm is given in \texttt{Algorithm \ref{alg:IMC}} below. The general procedure involves simulating the emission, absorption, and scattering of particles advanced along discrete time-steps. The absorbed/deposited energy in each time-step is used to calculate updated material temperatures which change the values of temperature-dependent quantities and the black-body distribution used to emit particles at the next time-step. As a Monte Carlo simulation, the interactions of individual particles are modeled probabilistically with the occurrence and outcomes of events such as scattering collisions being determined by random sampling.  
Quantities such as deposited energy are calculated by tallying the many energy deposition events that occur during a time-step. The challenges of accurately summing large lists of floating-point values (particularly in reduced-precision) motivate the techniques discussed in sub-section \ref{sec:SumAlgs}. More detailed descriptions of the IMC algorithmic procedure are given in \cite{FLECK1971313, 40IMC}.

\begin{algorithm}[H]
\caption{Implicit Monte Carlo Simulation}\label{alg:IMC}
\begin{algorithmic}[1]
\Function{IMC}{$\text{Input Parameters}$}
    \State{Create spatial mesh and array quantities}
    \For{$t \gets 0$ to $t_{end}$}
        \State{Update time-dependent quantities: $\beta,\sigma, f$}
        \State{Source new particles using Black-body temperature distribution}
        \For{Particle in Particles}
            \While {Particle history is active}
            \State{Calculate the boundary, collision, and census distances and choose the minimum}
            \State{Calculate new particle energy and deposit lost energy to material}
            \State{Advance particle position, time, and energy}
            \If{Particle energy is zero or domain boundary crossed}
                \State{Terminate particle}
            \EndIf
            \If{Census time reached}
            \State{Finish particle history and save properties to census}
            \EndIf
            \If{Particle experienced collision}
            \State{Calculate new direction (and frequency if not using grey approximation)}
            \EndIf
            \EndWhile            
        \EndFor
        \State{Remove terminated particles from census list}
        \State{Calculate end of time-step quantities: deposited energy and material/radiation energy densities}
        \State{Calculate new temperatures and advance to next time-step}
    \EndFor
    \State \Return{Outputs}
\EndFunction
\end{algorithmic}
\end{algorithm}

\subsection{Floating-Point Numbers}
Floating-point numbers are a subset of the real numbers $\mathbb{R}$ used for computer calculations. Since floating-point data-types are implemented using a fixed number of bits, they can only represent a discrete, finite subset of the full continuum of real numbers. Various arithmetic operations can be performed with floating-point numbers, however the results of floating-point arithmetic will typically not match exactly to the real number arithmetic. Calculated values typically have to be rounded to the nearest representable floating-point value.
In this work we consider floating-point arithmetic as implemented by the IEEE 754 standard \cite{IEE754}. AN IEEE 754 binary float can be denoted with a single sign bit $s$, $e$ exponent bits, and $d$ significand bits.
The actual binary numbers storing the exponent and significand are denoted as $E$ and $D$ respectively. The exponent is biased by an offset amount $2^{e-1}-1$ to allow for numbers both smaller and larger than $1$ to be represented. The significand is written as $1$ plus a fractional amount created by dividing the significand bits by $2^d$. The first bits indexed from $0$ to $d-1$ are occupied by the significand, the exponent from bits $d-1$ to $d-1+e$ and the sign bit is last at index $d+e$. A binary number can be written using the notation $b_i$ to denote the bit with index $i$ as:  
\begin{equation}
    (-1)^{b_{d+e}} \times 2^{(b_{d+e-1}b_{d+e-2}...b_{d-1})_2-(2^{e-1}-1)} \times(1.b_{d-1}b_{d-2}...b_0)_2,
\end{equation}
where the floating-point value can be expressed as:
\begin{equation}
    \text{fpv} = (-1)^s \times 2^{(E-(2^{e-1}-1))} \times (1+\frac{D}{2^d}),
\end{equation}
in which:
\begin{align}
    E = \sum_{i=0}^{e-1}b_{d+i}2^{i+1} \in \{1,2,...,(2^{e}-1)-1\}, \\
    D = \sum_{i=1}^{d} b_{d-i}2^{-i} \in \{1, 1+2^{-d},...,2-2^{-d}\}, \\
    s = b_{d+e-1}
\end{align}

Beyond the standard range of representable values, IEEE 754 floating-points have a few special characters, namely Inf's and NaN's \cite{GOLDBERGD1991Wecs}. The Inf and -Inf flags represent infinite values $\pm \infty$ and may arise when trying to calculate quantities such as $\pm1/0$ or $\pm \log(0)$. The NaN is used for calculations with indeterminate results such as $0/0$. There are also provisions made for denormalized numbers which allow for values smaller than the smallest normal floating-point to still be represented. This is done by allowing the leading bit of the significand to be zero, instead of one as it otherwise always is. Denormalized numbers are primarily used to avoid issues with two floats $x$ and $y$, where $x \neq y$ but whose difference $x-y$ is smaller than the smallest normal floating-point value. Upon evaluating $x - y$, the result would underflow to zero, which could lead to falsely identifying $x = y$, causing difficulty inidentifying bugs in code. 

The IEEE 754 standard defines several different binary data types including binary16/half, binary32/single, binary64/double, and binary128/quadruple. A comparison of relevant parameters for half, single, and double-precision variables is shown in Table \ref{tab:binaryfloats} below:
\begin{table}[H]
\centering
\resizebox{\textwidth}{!}{
\begin{tabular}{|c|c|c|c|}
\hline
\textbf{Base 2 Floats} & \textbf{Half-Precision} & \textbf{Single-Precision} & \textbf{Double-Precision} \\ 
\hline
\textbf{Number of Bits} & 16 & 32 & 64 \\ 
\hline
\textbf{Smallest Normal Positive Value} & $6.10 \times 10^{-5}$ & $1.18 \times 10^{-38}$ & $2.23 \times 10^{-308}$ \\ \hline
\textbf{Largest Value} & $65504$ & $3.40 \times 10^{38}$ & $1.80 \times 10^{308}$ \\ \hline
\textbf{Significand Digits (Binary/Decimal)} & $11 / 3.31$ & $24 / 7.22$ & $53 / 15.95$ \\ \hline
\end{tabular}
}
\caption{Binary Floating-Point Precision Comparison}
\label{tab:binaryfloats}
\end{table}
This table shows that the range of half-precision floats is significantly smaller than those of its single- and double-precision
counterparts. This limitation motivates the scaling techniques discussed in Section \ref{SM}.

The IEEE 754 standard is designed to guarantee commutativity (with the exception of calculations that result in special values like NaN or Inf). However, other algebraic properties of real numbers like associativity and distributivity do not necessarily hold. A summary of the validity of these properties in the real number field $\mathbb{R}$ and floating-point arithmetic is given in Table \ref{tab:algproperties}.
\begin{table}[H]
\centering
\resizebox{\textwidth}{!}{
\begin{tabular}{|l|c|c|}
\hline
\textbf{Algebraic Property} & \textbf{Real Numbers} $\mathbb{R}$ & \textbf{Floating-Point Numbers} \\ \hline
Associativity $a + (b + c) = (a + b) + c$   & $\checkmark$ Holds     & $\times$ Does not hold     \\ \hline
Commutativity $a \cdot b = b \cdot a$              & $\checkmark$  Holds     & $\checkmark$ Generally holds \\ \hline
Distributivity $a \cdot (b + c) = a \cdot b + a \cdot c$ & $\checkmark$ Holds     & $\times$ Does not hold     \\ \hline
\end{tabular}
}
\caption{Algebraic Properties: Real Numbers vs Floating-Point Numbers}
\label{tab:algproperties}
\end{table}

\section{Arithmetic Techniques}
\label{sec:AT}
\subsection{Summation Algorithms} \label{sec:SumAlgs}
Monte Carlo methods sum the results of repeated random sampling to form tallies which can be used to calculate various quantities of interest. We can express a generic tally $S_n$ formed from the sum of $N$ values $x_i$ as:
\begin{equation}
    S_n = \sum_{i=1}^N x_i
\end{equation}
The order in which variables are added together does not matter when working in the field of real numbers $\mathbb{R}$, due to the associative property of real numbers. 
However, the choice of numerical summation algorithm has an important effect on both accuracy and performance when calculating floating-point sums. Three summations methods: naive summation, pairwise summation, and compensated (Kahan) summation will be compared and contrasted below \cite{FPAccuracy93, KahanSum}.
\subsubsection{Naive Summation}
Naive summation is the simplest summation method that just sequentially adds terms to an accumulating sum. One advantage of this method is that no \textit{a priori} knowledge of the values to be summed is required, allowing the method to be easily implemented serially or in parallel. This property is useful for tallies in Monte Carlo codes where new values to be added are sampled stochastically. The most significant downside to this approach is the potential for catastrophic round-off error when the value to be added to a sum is smaller than the difference between the floating-point sum and the next representable value. The sum will remain unchanged if the additional values to be added are of equal or lesser magnitude. This error mechanism can lead to extremely large numerical errors in situations where many small values must be added together. The worst case round-off error for naive summation grows as $\mathcal{O}(\varepsilon n)$ and occurs when the round-off error of each addition has the same sign \cite{FPAccuracy93}. In cases where round-off errors can have opposing signs, the average RMS round-off error will be $\mathcal{O}(\varepsilon\sqrt{n})$.

\subsubsection{Pairwise Summation}
Pairwise summation takes a list of numbers and recursively splits it into halves until individual pairs are obtained. These pairs are then added and then the partial sums are summed together in multiple stages until the whole sum has been evaluated. For a list of numbers $x_i$ with $i \in [1,n]$ to be summed, the approach takes the form:
\begin{gather}
    S_n =\sum_{i=1}^n x_i \\
    = \sum_{i=1}^{n/2} x_i \ + \sum_{i=n/2+1}^n x_i \\
    \vdots \notag \\
    =\sum_{i=1}^2 x_i + \sum_{i=3}^4 x_i \ + \ ... \ +\sum_{i=n-3}^{n-2} x_i \ + \sum_{i=n-1}^n x_i
\end{gather}
This approach is significantly more accurate when summing a large number of values in reduced precision, as the magnitude of values added in each stage of summation are much more likely to be similar, greatly mitigating round-off error. A disadvantage for Monte Carlo applications is that pairwise summation requires the list of values to be known ahead of time. This requires saving arrays of sampled values to be added to a tally, increasing memory requirements. A hybrid approach may be developed that only saves lists up to a certain size before evaluating the sum, and then computing that partial sum and creating a new list for additional values to be added. This allows for much of the accuracy gains from pairwise summation to be retained while reducing storage requirements when summing up large numbers of values that are not all known ahead of time. The choice of list size must be carefully chosen based on hardware requirements and numerical accuracy considerations specific to the given problem. Pairwise summation has less round-off error than naive summation with a worst-case growth $\mathcal{O}(\varepsilon \log n)$ and average RMS error $\mathcal{O}(\varepsilon \sqrt{\log n})$ \cite{FPAccuracy93}. 

\subsubsection{Kahan Summation}
Kahan summation is similar to naive summation, but keeps an additional compensation term for low-order bits that would otherwise be lost to rounding-errors \cite{KahanSum}. The compensation term is then added back to the growing sum to prevent its value from being lost. The error growth of Kahan summation is technically $\mathcal{O}(n \varepsilon^2)$, but since floating-point values are rounded to a precision $\varepsilon$, the error term $n \varepsilon^2$ will round to zero, unless $n \gtrapprox 1/\varepsilon$. Practically, the error will be $O(\varepsilon)$ for sums where $n < 1/\varepsilon$. For double precision (binary64), $1/\varepsilon_{64} \approx 1 \times 10^{16}$, for single precision (binary32), it is $1/\varepsilon_{32} \approx 1.8 \times 10^7$, and for half precision (binary16), it is $1/\varepsilon_{16} \approx 2.05 \times 10^3$. So, Kahan summation in half-precision can be less accurate than pairwise summation for larger $n$ values. A practical example of where this might occur is in Monte Carlo simulations of particles traversing highly collisional (optically thick) media where many small events are tallied. For smaller $n$ values, Kahan summation is the most accurate summation algorithm, although it requires roughly four times as many operations as naive and pairwise summation. A comparison of the algorithmic complexity and accuracy of the three summation algorithms is given in Table \ref{tab:SumAlgs}.

\begin{table}[H]
    \centering
    \resizebox{\textwidth}{!}{%
    \begin{tabular}{|c|c|c|c|}
         \hline
         \textbf{Summation Algorithm} & \textbf{Arithmetic Operations} & \textbf{Worst-Case Error} & \textbf{Average RMS Error} \\
         \hline
         Naive & $ \sim n$ (parallelizable)  & $\mathcal{O}(\varepsilon n)$ & $\mathcal{O}(\varepsilon\sqrt{n})$ \\
         \hline
         Pairwise & $ \sim n $ (parallelizable) &$\mathcal{O}(\varepsilon \log n)$& $\mathcal{O}( \varepsilon \sqrt{\log n})$ \\
         \hline
         Kahan (Compensated) & $ \sim4n$ (serial) & $\mathcal{O}(n\varepsilon^2) \approx \mathcal{O}(\varepsilon) $& $\mathcal{O}(n\varepsilon^2) \approx \mathcal{O}(\varepsilon)$ \\
         \hline 
    \end{tabular}
    }
    \caption{Summation Algorithms Comparison}
    \label{tab:SumAlgs}
\end{table} 

\subsection{Accurate Evaluation of $\left(1-\exp{(x)}\right)$} \label{sec:expm1}
A common variance reduction technique for Implicit Monte Carlo simulations is implicit capture. If a particle with energy $E$ experiences an implicit capture event, it will decrease in energy to  $E'= E \exp(-f\sigma_a d)$ and deposit energy $E_{dep} = E(1-\exp(-f \sigma d))$ in the material. In situations where $\exp(x)$ is close to $1$ $(\geq 0.999)$ the expression $1-\exp(x)$ can round to zero. This is problematic for IMC simulations, resulting in particles not depositing energy, which can lead to underheating in the material. One approach to address this issue is to use the Taylor series expansion $1 - \exp(x) \approx 1 - (1+ x + \frac{x^2}{2} + \frac{x^3}{6} + \cdots) = -x - \frac{x^2}{2} -\frac{x^3}{6} - \cdots$.
Another more robust approach is to evaluate the expression $\exp(x)-1$ using an \texttt{expm1(x)} function, which is designed to reduce round-off errors when $\exp(x)$ is close to one \cite{expm1}. We can obtain its complement $1-\exp(x)$ as -expm1(x) since the argument $x<0$ and $\exp(x) \leq 1$ for implicit capture calculations. A key take-away from this is that other floating-point mathematical expressions used in simulations should be checked to confirm that they evaluate as expected for the given range of input values. Doing so can help avoid the occurrence of many potentially difficult-to-diagnose error modes.

\subsection{Rearranging the Order of Operations}
In some instances, algorithms that change the order of operations of products can be used to prevent the occurrence of under/overflow in intermediate calculations for cases where the final result fits within the representable range of a floating-point datatype. A simple example application is the product of three numbers in floating-point arithmetic, two large ($l_1$ and $l_2$) and one small ($s$), whose product is representable. If the product is evaluated by multiplying the two large numbers together first, such as $l_1\times l_2$ the result could overflow as the intermediate calculation is too large to be representable. However, if a small and large number are multiplied together first, such as $l_1 \times s$ or $l_2 \times s$, the intermediate result will be representable, safely allowing the other large number to be multiplied with it. The order of operations algorithm orders all multiplicands by size and pairs the largest and smallest values, the second largest and smallest, and so on. The product is then evaluated in the order of these arranged pairs. The algorithm can also rescale the product if the result would not otherwise be representable, or the user wishes to change the scale for use in other calculations. Given a list of different scales, the algorithm will sort them by size and return the largest scaled product that does not overflow. It takes as input a list of multiplicands and a list of potential scale factors, and returns a product and its scale factor as outputs. 

A pseudo-code  implementation is shown in \texttt{Algorithm \ref{alg:moor}}. This algorithm can also be used without scale factors (equivalently setting a single scale of size unity) to rearrange products appearing in more complicated algebraic expressions. A practical example is the reordering of the product $\alpha c \sigma_{a,p}^n \Delta t \beta^n$ present in the denominator of the Fleck factor to prevent it from overflowing. Without this reordering, the product in the denominator can in some cases equal \texttt{Inf} resulting in a division of $1/\infty \rightarrow 0$ that would cause the Fleck factor to equal zero. A null Fleck factor is nonphysical and prevents the deposition of particle energy in materials, causing the IMC simulation to stagnate.

\begin{algorithm}[H]
\caption{Multiplicative Order of Operations Rearranger}\label{alg:moor}
\begin{algorithmic}[1]
\Function{Multiplication Rearranger}{$\text{multiplicands}, \text{scales}$}
    \For{$j \gets 1$ to length(scales)}  \Comment{Loop through potential scale factors}
    \State{product $\gets 1.0$} \Comment{Initialize product to 1.0}
    \State{sorted $\gets$ sort(multiplicands, scales[j])} \Comment{Sort list of values to multiply}
    \For{$i \gets 1$ to length(sorted)/2}
    \State{product $\gets$ product $\times$ sorted[i] $\times$ sorted[end-i-1]} \Comment{Multiply the smallest and largest elements and so on}
    \EndFor
    \If{length(sorted) is odd}
    \State{product $\gets$ product $\times$ sorted[length(sorted)/2 +1]} \Comment{Multiply middle element if odd number of elements}
    \EndIf
    \If{(product != Nan and product != Inf)}
    \State \Return (product, scales[j]) 
    \EndIf
    \EndFor
    \State{print(Product not representable)}
    \State \Return{(0.0,0.0)} \Comment{Return default values if product not representable}
    
\EndFunction
\end{algorithmic}
\end{algorithm}

\section{Scaling Methods}
\label{SM}
The primary goal of scaling methods is to numerically resize floating-point variables such that calculations performed with them fit within the representable range of the floating-point datatype. This is especially important when working with the condensed range available to reduced-precision floats. Scaling methods can be static, as in changing the base units of physical quantities such as distance or time. They can also be dynamic with scale factors that vary throughout a simulation, such as in the case of reciprocal scale factors used to normalize variables. Caution must be taken to either undo scaling or note the change in scale when interpreting results calculated with scaled variables. The optimal choice of scale factors is typically not apparent \textit{a priori} and will often require empirically derived knowledge of the numerics of a given problem. Using scale factors that are multiples of two can help minimize round-off errors in binary floats, as multiplying them with a binary float will just shift bits in the exponent and not change the fractional part in the mantissa bits. However, some strategies and algorithmic approaches for scaling are discussed below. 

\subsection{Distance Scaling}
A prerequisite requirement for effective distance scaling is the use of relative position tracking \cite{alexlong24}. The relative position tracking method redefines the position of particles relative to the boundaries of their current spatial cell, instead of measuring against a common global datum/origin. Converting between absolute and relative positions is straightforward when using a Cartesian coordinate system. Starting from the origin, we can index the number of cells between it and any given location as $n_x$, $n_y$, and $n_z$ in the $x$, $y$, and $z$ directions respectively. If the widths of cells in each direction are stored in arrays $dx$, $dy$, and $dz$, we can convert between the absolute position $[x_{abs}, y_{abs}, z_{abs}]$ and relative position $[x_{rel}, y_{rel}, z_{rel}]$ as:
\begin{equation}
    [x_{rel},y_{rel},z_{rel}] = [x_{abs},y_{abs}, z_{abs}] - [\sum_i^{n_x} dx_i, \sum_j^{n_y}dy_j, \sum_k^{n_z}dz_k]
\end{equation}
There are several advantages to using relative instead of absolute positions when working in reduced-precision floating-point arithmetic. The first is that position values will always be limited by the smaller cell sizes rather than by the global dimensions of the problem domain. This allows for more accurate tracking of particle position changes due to fewer round-off errors since changes in position are numerically closer in magnitude to the cell size. In particular, it prevents particles in optically thick media far from the origin from becoming ``stuck" when the distance to travel is smaller than the spacing to the next representable floating-point value. Relative position tracking also allows distances to be numerically rescaled on an individual cell-wise basis. A similar strategy can be developed for relative time-step tracking, which resets the time variable used to track particle events at the end of each time-step. This allows for more accurate tracking of short duration events, particularly those that occur at later times from the start of the simulation. The relative time-tracking concept can also be applied to the global simulation time variable, by using an additional time-step index variable to track the number of time-steps that have passed.

The distance scaling method recognizes that the exponential attenuation of particles through a medium is governed by the product of the macroscopic cross-section $\sigma$ and distance $d$. By introducing a distance scale factor $S_d$, we can define a reciprocally scaled macroscopic cross-section $\sigma' = \sigma/S_d$ and a scaled distance $d' = S_d d$:
\begin{equation}
    \sigma d = \frac{\sigma}{S_d} S_d d = \sigma ' d'
\end{equation}
If the distances traversed by particles in a cell are scaled up by $S_d$, then the macroscopic cross-section $\sigma$ can be replaced with $\sigma'$ without changing the results of calculations for things like collision distance or energy deposition. This is useful in situations where the unscaled macroscopic cross-section $\sigma$ is too large to represent in reduced-precision floating-point arithmetic. This approach can effectively be thought of as changing the units used to measure distance inside of a given spatial cell. 

\subsection{Multiple Energy Scaling}
An extremely valuable technique for thermal radiation transport problems is the use of multiple energy scales. The radiative transfer equation's nonlinear $T^4$  dependence on temperature and other highly variable quantities such as opacity or cell size means that energy can range over many orders of magnitude. In reduced-precision floating-point calculations, this issue is especially problematic as sourcing energy can either be too large or too small to represent. It also means that a single energy scale factor will likely be inadequate and different scales will be needed across both space and time. In the multiple energy scale approach,  an energy scale factor is assigned to each spatial cell during sourcing to ensure that its energy is in a representable range. Each particle sourced into a cell during the time-step will have a scaled energy weight and be assigned the energy scale factor as an additional property. Particles will undergo transport as usual using scaled energies, with the exception that energy depositions will be made into separate arrays corresponding to individual scale factors. 

This approach can be more precisely formulated by considering a spatial mesh comprised of $N$ cells with sourcing energies $E_1, \ldots, E_N$. The energy in each cell can be scaled using one of $M$ energy scales where $M \in \mathbb{Z} :1 \leq M \leq N$. The scaled cell energies are denoted as $E_i'= S_jE_j$ where each energy scale $j \in [1,M]$ has $j_m$ particles sourced into it. At each time-step, separate transport loops will be executed for each energy scale with corresponding scaled energy deposition arrays, census energy tallies, and lost energy tallies. This approach has been shown~\cite{butson2025} to conserve energy, so long as energy values are appropriately unscaled to the same common units when tallying the redistribution of energy between the radiation field and the material at the end of time-steps.

Multiple energy scaling is particularly useful in situations where the total magnitude of the sourcing energy is low, but the number of particles to be sourced is large. In such situations, the energy allotted to each particle may be too numerically small to represent causing an underflow to zero. This will result in underheating and unphysical cooling because sourced energy is not assigned to transported particles. Using large scale factors can allow for these lower-energy particles to be numerically representable and also allow for smaller implicit capture energy depositions to still be accurately recorded. To successfully implement the multiple energy scaling technique, the automatic order of operations algorithm should also be used to calculate cell-wise sourcing energies and to select the largest workable energy scale from a user-provided list. This method is fairly robust; however, some degree of experimentation is required to determine suitable combinations of scale factors for specific problems. 

\section{Algorithmic Complexity}
While techniques such as pairwise summation and the order of operations rearranging algorithm improve the accuracy of reduced-precision computations, they do so at the expense of increased algorithmic complexity. If one's goal for using reduced-precision floats is to obtain an increase in computational performance, the accuracy vs. performance trade-off must be considered when choosing which techniques to implement. The increased memory requirement of the pairwise summation method is largely dependent on the number of energy deposition events in any given time step. For a grid with $N_x$ cells and $N_E$ energy deposition events in a given time step, a standard naive summation approach will deposit the $N_E$ events into elements of the $N_x$ array by adding each the energy associated with each deposition event to an existing energy value stored in the appropriate cell. Pairwise summation requires all $N_E$ events to be individually distributed and saved to a unique element (not added to a cell energy sum) in one of the $N_x$ energy deposition vectors which can range in length from $0$ to $N_E$ elements. Accordingly, the storage requirement for naive summation is $\mathcal{O}(N_x)$ elements whereas for pairwise summation it is $\mathcal{O}(N_E)$ where $N_E \gg N_x$ in most IMC simulations. When simulating optically thick media where $N_E$ is especially large, it can be advantageous to set a limit $N_{E,S}$ on the number of individual energy depositions that can be stored in each vector before they are accumulated. Doing so will change the overall energy deposition memory requirement to $\mathcal{O}(N_{E,S} N_x)$, which is reduced as long as $N_{E,S} < \lfloor \frac{N_E}{N_x} \rfloor$. 

The order of operations algorithm requires more operations than just directly evaluating a product. Its algorithmic complexity is estimated by decomposing it into sorting and multiplying operations. We assume the use of a quicksort algorithm with average complexity $\mathcal{O}(n \log n)$ \cite{introalgs}. Multiplying the sorted numbers will take $\mathcal{O}(n)$ operations and this whole procedure will be repeated $m$ times, where $m$ is the number of scaling factors attempted before a valid output is returned. Combining this information, we evaluate the average algorithmic complexity of the order of operations rearranging algorithm as:
\begin{equation}
    T_{avg}(m,n) = \underset{scaling \ attempts}{\underbrace{m}} \times \quad (\underset{sorting}{\underbrace{\mathcal{O}(n \log n)}} \quad + \underset{multiplies}{\underbrace{\mathcal{O}(n)}}) \approx \mathcal{O}(m n \log n)
\end{equation}
In the worst case, the quicksort algorithm has complexity $\mathcal{O}(n^2)$ which means the algorithmic complexity of the order of operations rearranging algorithm becomes:
\begin{equation}
    T_{worst}(m,n) = \underset{scaling \ attempts}{\underbrace{m}} \times \quad (\underset{sorting}{\underbrace{\mathcal{O}(n^2)}} \quad + \underset{multiplies}{\underbrace{\mathcal{O}(n)}}) \approx \mathcal{O}(m n^2)
\end{equation}

\section{Random Walk Acceleration}
Random walk acceleration is a technique that replaces a number of short-range collisions in a Monte Carlo simulation with a single longer random walk stride. This is advantageous in optically thick media where many collisions occur per unit distance. Each collision makes little contribution to problem tallies, yet every one must be computed to accurately model the problem. A single aggregate random walk can model the same (or very nearly the same) physics with significantly less computational effort. An additional potential advantage of random walks for reduced-precision floating-point implementations is that many smaller interaction collisions are combined, avoiding the need to apply scaling techniques to accurately model the shorter collision events. The approach we consider is a diffusion model developed for IMC by Fleck and Canfield \cite{FleckJ.A1984Arwp}.
This random walk approach can be applied if the following criteria hold: 
\begin{align}
    R_0 > \lambda, \\
    d_{col} = \frac{|\log \xi|}{\sigma_s} < R_0
\end{align}
where $R_0$ is the radius of a random walk sphere (the distance between the particle and the nearest cell boundary) and $\lambda$ is the mean free path of the particle. These two criteria help ensure that the random walk sphere is large enough to encompass the distance a particle might travel after several collisions or a single random walk. A standard IMC transport approach is used if the criteria are not met. 
The random walk approach uses a diffusion equation to compute the probability density function $\psi$ of a particle's spatial distribution over time:
\begin{align}
    \frac{\partial \psi}{\partial t} = D \nabla^2 \psi, \\
    \psi(\textbf{r},0) = \delta(\textbf{r}), \\
    \psi(R_0,t) = 0    
\end{align}
The solution of the diffusion equation subject to the given initial and boundary conditions is:
\begin{equation}
    \psi(r,t) = \frac{1}{2R^2_0} \sum_{n=1}^\infty \Big(\frac{n}{r}\Big) \exp\Big(-\Big(\frac{\pi n}{R_0}\Big)^2 Dt \Big) \sin \Big(\frac{n \pi r}{R_0}\Big)
\end{equation}
The probability that a particle remains inside the random walk sphere for a time $t$ is:
\begin{align}
    P_R(t) =  \int_0^{2\pi}\int_0^\pi\int_0^{R_0} \psi(r,t) r^2 \sin(\theta) dr d\theta d\phi = \\
    4 \pi \int_0^{R_0} \Big( \frac{1}{2R^2_0} \sum_{n=1}^\infty \Big(\frac{n}{r}\Big) \exp \Big(-\Big(\frac{\pi n}{R_0}\Big)^2 Dt \Big) \sin \Big(\frac{n \pi r}{R_0}\Big)r^2 dr \Big) \\
    = -2 \sum_{n=1}^{\infty} (-1)^n \exp(-\frac{\pi n}{R_0}^2 Dt)
\end{align} 
The probability of the particle terminating its random walk by arriving at the sphere surface is:
\begin{equation}
    P_T(t) = 1 - P_R(t) = 1 + 2 \sum_{n=1}^{\infty} (-1)^n \exp(-\frac{\pi n}{R_0}^2 Dt)
\end{equation}
In the Monte Carlo simulation, we evaluate whether a particle has reached the sphere surface before the census time $t_{cen}$. If the particle reaches the surface, the particle's flight will be terminated; if not, the particle will continue moving in the next time-step. A uniform random variable $\xi \in (0,1]$ is sampled, and if $0 < \xi \leq P_t(t_{cen})$ then the particle reaches the surface. Otherwise, if $P_T(t_{cen}) < \xi \leq 1$ then the particle remains in the sphere at the census time. If the particle leaves the sphere, the actual exit time $t_T$ is computed by inverting the CPDF $P_T(t_T) = \xi$. A lookup table of $P_T(t)$ values that can be interpolated is recommended instead of repeatedly computing the series solution whenever a random walk occurs. It is ideal to generate the table using a range of exponential coefficient values $a = \frac{Dt}{R^2}$ as this allows for time $t$ and sphere radius $R$ to be computed independently. Practically, the infinite series can be evaluated with around 100 terms in floating-point arithmetic without any appreciable loss in accuracy/precision from the neglected higher terms. The lookup table can be computed once and saved to avoid redundant computational expense. If the particle reaches the surface of the sphere, it advances in time by an amount $t_T$, receives a new position uniformly sampled on the surface of the sphere $r = R_0$, and its direction is sampled from a cosine distribution relative to normal to the surface of the sphere. However, if the particle reaches census first it will have its time advanced by an amount $t_{cen}$, scatter isotropically, and have a new position sampled from a sphere $r = R_1$, which is computed from
\begin{align}
   P_{R_1}(t) =  \int_0^{2\pi}\int_0^\pi\int_0^{R_1} \psi(r,t) r^2 \sin(\theta) dr d\theta d\phi \psi(r,t_{cen})rdr = \xi_2 P_R(t_{cen}) \\
   \rightarrow -2 \sum_{n=1}^{\infty} (-1)^n \exp(-\frac{\pi n}{R_1}^2 Dt) = \xi_2 P_R(t_{cen})
\end{align}
where $\xi_2$ is a new uniform random variable to be sampled. A lookup table of $P_R$ using the exponential coefficient $a$ should be used to solve for $R_1 = \sqrt{\frac{a}{Dt_{cen}}}$. In practice, a random walk transport step is more computationally expensive than a standard transport step. To achieve a speed-up in runtime, the total number of transport steps must be reduced by a factor greater than the increase in average transport step runtime. This is often the case for problems that are entirely optically thick, but may not be for problems with both optically thick and thin regions. 

\section{Results}

\subsection{Su Olson Benchmark}
The Su Olson benchmark problem models non-equilibrium radiative transfer in an isotropically scattering medium \cite{SU19971035}. A source of width $x_0$ is active for a time $\tau_0$ around the origin of a cold, homogeneous, semi-infinite medium. The radiative transfer and material energy balance equations are linearized by assuming a heat capacity of the form $c_v = \alpha T^3$. These equations are analytically solvable and can be written in the following scaled form:
\begin{equation}
    \big(\varepsilon \frac{\partial}{\partial \tau} + \mu \frac{\partial}{\partial x} +1 \big) U(x,\mu,\tau )= \frac{c_a}{2}V(x,\tau) + \frac{c_s}{2} W(x,\tau) + Q(x,\mu,\tau) 
\end{equation}
\begin{equation}
    \frac{\partial V(x,\tau)}{\partial \tau} = c_a [W(x,\tau) - V(x,\tau)]
\end{equation}
with
\begin{align}
    U(x,\mu,\tau) = \frac{I(z,\mu,t)}{a T_{H}^4}, \quad V(x,\tau) = \Bigg[\frac{T(z,t)}{T_H}\Bigg]^4, \\
    W(x,\tau) = \int_{-1}^{1} d\mu U(x,\mu,\tau), \quad Q(x,\mu,\tau) = \frac{S(z,\mu,t)}{a T_{H}^4}.
\end{align}
where the quantities $U$, $W$, $V$, and $Q$ are normalized radiation intensity, radiation energy density, material energy density, and radiation source, respectively. The absorbing and scattering ratios are $c_a$ and $c_s$, and the scaled position and time are $x$ and $\tau$. The reference temperature $T_h$ is normalized along with the radiation constant $a$ and wave-speed $c$ to a value of $1$ in scaled units. The parameter $\epsilon = \frac{4 a}{\alpha}$ is also set equal to $1$.

\begin{figure}[H]
    \centering
    \includegraphics[width=0.8\linewidth]{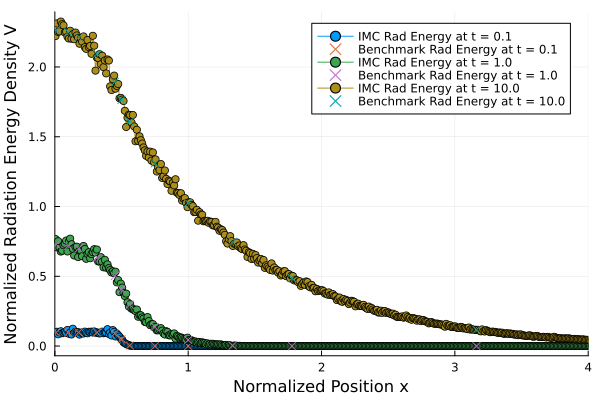}
    \caption{Su-Olson Problem Double-Precision Radiation Energy Density}
    \label{fig:radeng64}
\end{figure}

\begin{figure}[H]
    \centering
    \includegraphics[width=0.8\linewidth]{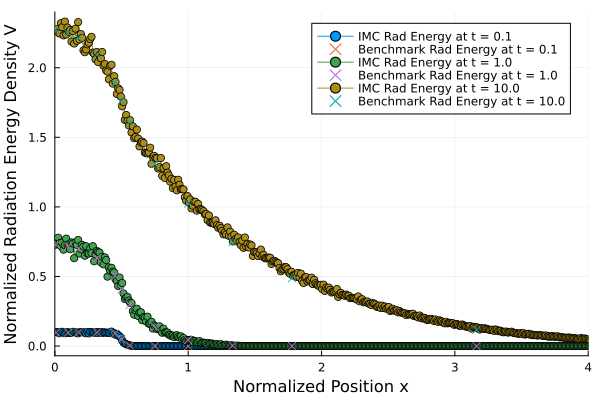}
    \caption{Su-Olson Problem Half-Precision Radiation Energy Density}
    \label{fig:radeng16}
\end{figure}

\begin{figure}[H]
    \centering
    \includegraphics[width=0.8\linewidth]{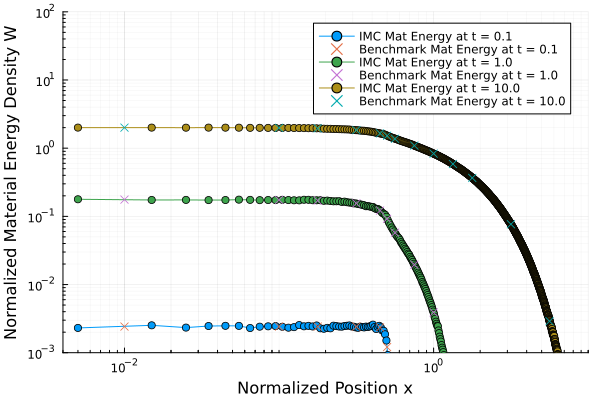}
    \caption{Su-Olson Problem Double-Precision Material Energy Density}
    \label{fig:mateng64}
\end{figure}

\begin{figure}[H]
    \centering
    \includegraphics[width=0.8\linewidth]{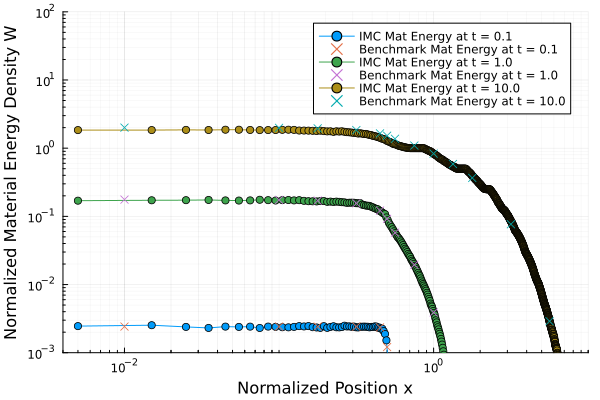}
    \caption{Su-Olson Problem Half-Precision Material Energy Density}
    \label{fig:mateng16}
\end{figure}

 The specific Su-Olson problem considered here has equal absorption and scattering ratios $c_a = c_s = 0.5$. It was solved using a standard IMC approach in both half and double float precision. A unit strength radiation source $Q_r$ with width $x_0 = 0.5$ and active duration $\tau_0 = 10$ is prescribed. IMC simulations were executed with 2000 time-steps of duration $\Delta t = 0.005$ to reach a maximum time of $\tau = 10$. This time-step was experimentally determined to be short enough to prevent overheating at earlier times. The spatial cells used have width $\Delta x = 0.01$ and the number of source particles introduced in each time-step is $N_S = 1000$. A constant energy scale factor of $S = 32768$ was used to produce the half-precision results. This scaling was required to prevent smaller energy depositions from underflowing,  underheating the material and undervaluing later material and radiation energy density values. Distance scaling was not used as the total opacity is already unity and spatial cell widths were chosen appropriately. A reflecting boundary condition is used at the left boundary and there is a vacuum boundary present at $x = 10$. This vacuum boundary distance was chosen to be sufficiently far away from the origin for the IMC results to match the benchmark without having to simulate the entire half-space $0<x<\infty$. Standard naive summation was used; pairwise summation yielded minimal improvements to accuracy in this problem compared to its increased runtime/memory usage. The Su-Olson benchmark analytic transport results are used for comparison with the results from the IMC simulation. The radiation energy density results for the double and half precision cases are shown in Figures \ref{fig:radeng64} and \ref{fig:radeng16}, while the double and half precision material energy density results are shown in Figures \ref{fig:mateng64} and \ref{fig:mateng16}.

 The IMC results for both double and half-precision cases converge to the material and radiation energy density benchmark solutions. The radiation energy density results increase over time as more and more radiation energy is introduced into the system from the radiation source and from black-body radiation emitted by the increasingly heated material. The material energy densities for both float precisions match the benchmark values closely, with the primary difference being some inconsistencies in the form of steps or piecewise-constant sections in the material energy density wave-front/tail at later times. These can be mainly attributed to round-off error that occurs when many small energy deposition events are accumulated. Regardless, there is still good agreement between the benchmark and IMC results in the tail towards the right of the slab. 

\subsection{Marshak Wave}
\begin{figure}[H]
    \centering
    \includegraphics[width=0.8\linewidth]{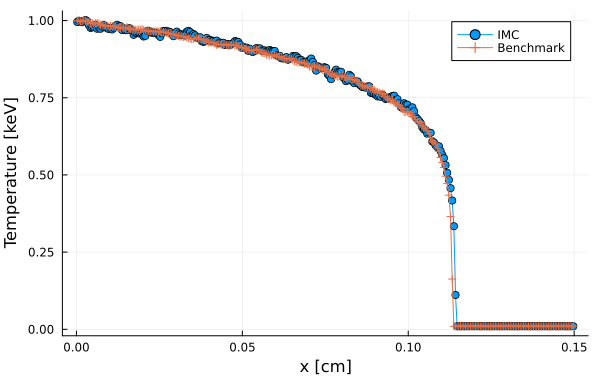}
    \caption{Marshak Wave Double-Precision at Time $t = 2.00$ Shakes}
    \label{fig:marshak64}
\end{figure}

\begin{figure}[H]
    \centering
    \includegraphics[width=0.8\linewidth]{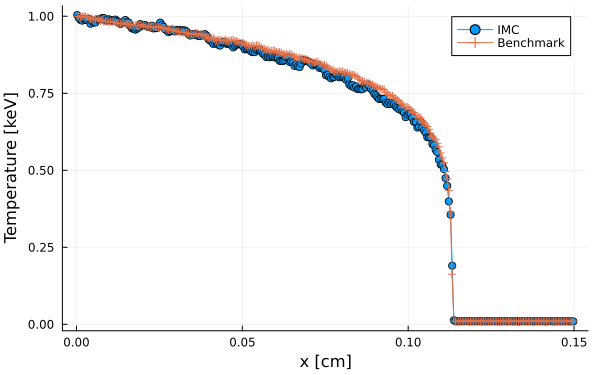}
    \caption{Marshak Wave Half-Precision at Time $t = 2.00$ Shakes}
    \label{fig:marshak16}
\end{figure}

A Marshak wave \cite{MarshakR.E.1958EoRo} is created by taking an initially cold homogeneous slab with a temperature dependent opacity $\sigma_a(T) = 1000/T^3 \text{ cm}^{-1}$ and applying a $1$ keV temperature source to its left boundary. The initial slab temperature of $T_{init}= 0.01$ keV  results in a very large starting opacity of $\sigma(0.01 \text{ keV}) = 1000/(0.01 \text{ keV})^3 = 1 \times 10^9 \text{ cm}^{-1}$. This opacity greatly exceeds the maximum representable value of a half-precision floating-point variable and requires distance scaling to be representable. A large distance scale factor $S_d = 50000$ was used to allow unscaled opacity values as large as $1 \times 10^9 \text{ cm}^{-1}$ to be represented in a scaled form in half-precision. A ramping time-step is used to more accurately capture the initial formation of the Marshak wave, while minimizing computational effort at times after the shock wave front has formed and is just propagating. This ramp scheme prescribes the time step $dt_n$ in step $n$ as
\begin{equation}
    dt_n = \min(k^{n}dt_0, dt_{max})
\end{equation}
The values used in this problem are initial time-step size $dt_0 = 1\times10^{-5}$ shakes, multiplicative factor $k = 1.01$, maximum time-step size $dt_{max} = 1\times10^{-2}$ shakes, and an end time of $2.0$ shakes. Many of the energy deposition events will be numerically small due to the large opacity and short time steps, which can cause underflow to occur and lead to underheating of the material. The opposite can occur at later times when larger time-steps are used where cold cells with corresponding very large opacities may lead to an overflow in the sourcing energy for those cells. To address both of these issues, we will use multiple energy scaling with the following list of base-2 scale factors: $[1024.0, 128.0, 8.0, 1.0,0.125, 0.015625]$. This includes factors larger and smaller than $1.0$ to avoid both overflow and underflow problems. They are also multiples of two to minimize rounding errors when scaling. The automatic order of operations rearranging algorithm was also utilized in reduced precision to prevent the product in the denominator of the Fleck factor ($\alpha c \sigma_{a,p}^n \Delta t \beta ^n$) from overflowing and causing the Fleck factor to round to zero. A null Fleck factor prevents energy from being deposited in cells in the implicit capture method, which makes the Marshak wave stagnant. We compare against benchmark results generated by the Kull IMC code developed by Lawrence Livermore National Laboratory \cite{Gentile1998}. Plots of the double and half-precision Marshak Waves compared against Kull IMC results are shown in Figures \ref{fig:marshak64} and \ref{fig:marshak16}. 

In both float-precisions the Marshak wave results match closely to the Kull IMC results, however the reduced-precision results have a slightly lower temperature than the benchmark near the wavefront/tail. This decreased temperature is attributable to round-off losses when accumulating numerous numerically small energy depositions over many time-steps. It should be noted that the half-precision Marshak wave results shown here use the typically less accurate $\exp(x)$ expression to evaluate energy depositions rather than the \texttt{expm1} method discussed in section \ref{sec:expm1}, as it leads to a better alignment of the wavefront with the double-precision and Kull IMC wavefronts. Using the \texttt{expm1} expression for this problems reduces the wave-speed and causes the reduced-precision wavefront to lag behind its expected position. This effect is most likely due to the \texttt{expm1} method depositing less energy into the material at earlier time-steps causing the material to remain colder and more optically thick, slowing the propagation of the wavefront. The Marshak wave problem highlights the efficacy of the distance scaling method in enabling the calculation of values that are otherwise too large to represent, and the difficulties of reduced-precision floating-point calculations evaluating in unexpected ways.

\subsection{Random Walk Acceleration}
To demonstrate the efficacy of the random walk method, we consider an optically thick infinite medium problem. This problem has a uniform temperature of $1$ keV, a constant opacity $\sigma = 1000 \text{ cm}^{-1}$ and a heat capacity $c_v = 1 \frac{\text{jk}}{\text{cm}^2\text{keV}}$. This problem was simulated in double, single, and half floating-point precision, with and without random walk acceleration. In each case, a uniform spatial cell width of $0.05 \text{ cm}$, a time step size $\Delta t = 0.0005 \text{ shakes}$, and $N_s = 10000$ particles was used. The runtime and number of transport iterations (passes through the Monte Carlo transport loop) for each test case are included in Table \ref{tab:randomwalks}.

\begin{table}[H]
    \centering
    \resizebox{\textwidth}{!}{%
    \begin{tabular}{|c|c|c|c|}
    \hline
        \textbf{Float-Precision} & \textbf{No Random Walk} & \textbf{Random Walk} & \textbf{Relative Change} \\ \hline
        Double & 824.97 seconds & 159.03 seconds & 5.19x faster \\
               & 791,452,153 iterations &  25,908,840 iterations & 30.54x fewer \\ \hline
        Single & 842.31 seconds & 170.01 seconds & 4.95x faster \\
               & 791,387,334 iterations &  25,919,523 iterations & 30.53x fewer \\ \hline      
        Half & 826.69 seconds & 166.76 seconds & 4.96x faster \\
               & 790,615,119 iterations &  26,556,294 iterations & 29.77x fewer \\ \hline  
    \end{tabular}
    }
    \caption{Performance Comparison With \& Without Random Walk Acceleration In Optically Thick Problem}
    \label{tab:randomwalks}
\end{table}
The random walk-accelerated cases experience about a 5x speed-up in runtime and about a 30x reduction in the number of transport iterations. Accordingly, we observe that a random walk transport step is approximately 6x slower than a traditional transport step in our implementation. While the random walk accelerated the run-time of this problem, it is not guaranteed to do so for all problems. The random walk method is most effective for problems that are optically thick and have sufficiently large spatial cells such that the reduction in the number of transport iterations will greatly outweigh the increased runtime of the random walk transport steps. Extending the random walk acceleration method to reduced-precision did not require additional modifications beyond the arithmetic manipulations and scaling methods previously discussed in this paper.

 
\subsection{Crooked Pipe}
Here, we consider a two-dimensional IMC benchmark, known as the crooked pipe problem~\cite{CrookedPipe}, consisting of a bent pipe geometry with an optically thin inner region surrounded by an optically thick outer medium. The problem begins with a global equilibrium temperature of $0.05$ keV, except for a hotter $0.5 $ keV surface source applied to the lower left pipe opening. Radiation from the temperature source flows easily through the first pipe segment, heating the optically thick wall segments visible to it. As these walls heat up, they emit more radiation which in turn travels through the pipe, heating up other wall segments with no direct view of the original source. Over time, the entire bent pipe geometry heats up, with a noticeable temperature gradient that extends from the source to the other end of the pipe. This problem is defined in a cylindrical 3D geometry, but due to rotational symmetry can be modeled in 2D cartesian coordinates using the length and radius of the pipe as $x$ and $y$ coordinates. Accordingly, there are vacuum boundary conditions on the left, right, and top boundaries and a reflecting boundary on the bottom boundary which runs along the centerline of the pipe. The optically thin material has an opacity $\sigma_a = 0.2 \text{ cm}^{-1}$ and heat capacity $c_v = 1\times10^{-3} \frac{\text{jk}}{\text{cm}^2\text{keV}}$, while the optically thick material has and opacity $\sigma_a = 2000 \text{ cm}^{-1}$ and heat capacity $c_v = 1\ \frac{\text{jk}}{\text{cm}^2\text{keV}}$. 

The specification of the crooked pipe problem involves a nonuniform mesh spacing in the $x$ and $y$ dimensions, with smaller cells around the pipe walls to better resolve temperature changes there. The simulations were performed using approximately $80,000$ particles per time-step for a total duration of $100$ shakes. A ramp time-step was used with an initial time-step size $dt_0 = 0.001$ shakes, a multiplicative factor $k=1.1$ and a maximum time-step size $dt_{max} = 0.1$ shakes. The double-precision simulation did not employ energy scaling, and energy deposition events were tallied using a standard naive summation approach. In contrast, the half-precision simulation required the multiple energy scaling approach and pairwise summation for energy depositions. The list of multiple energy scale factors used was: $[50000,25000,10000,5000,1000,100,10,1,0.1]$. The automatic order-of-operations algorithm was utilized to calculate the sourcing energy and select an appropriate energy scale for each cell. Temperature heatmaps at $t=10.0$ shakes produced by the double and half-precision simulations are shown in Figures \ref{fig:crookedpipe64} and \ref{fig:crookedpipe16} below.

\begin{figure}[H]
    \centering
    \includegraphics[width=0.8\linewidth]{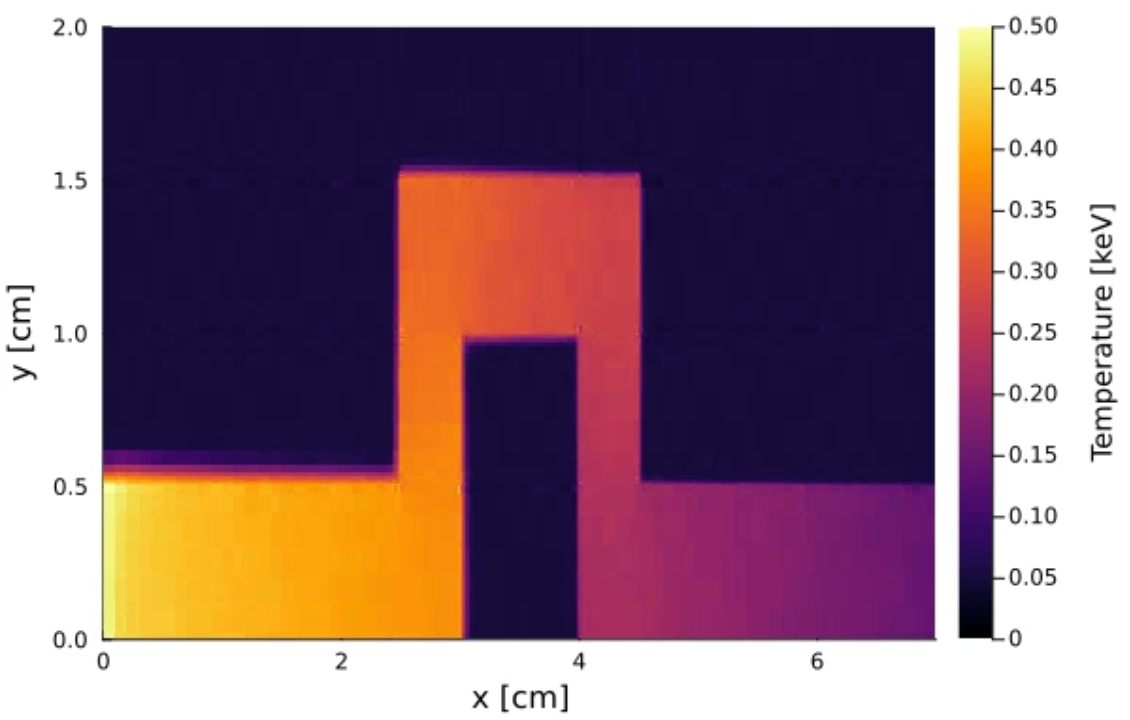}
    \caption{Crooked Pipe Double-Precision Heatmap at Time $t = 10.0$ Shakes}
    \label{fig:crookedpipe64}
\end{figure}

\begin{figure}[H]
    \centering
    \includegraphics[width=0.8\linewidth]{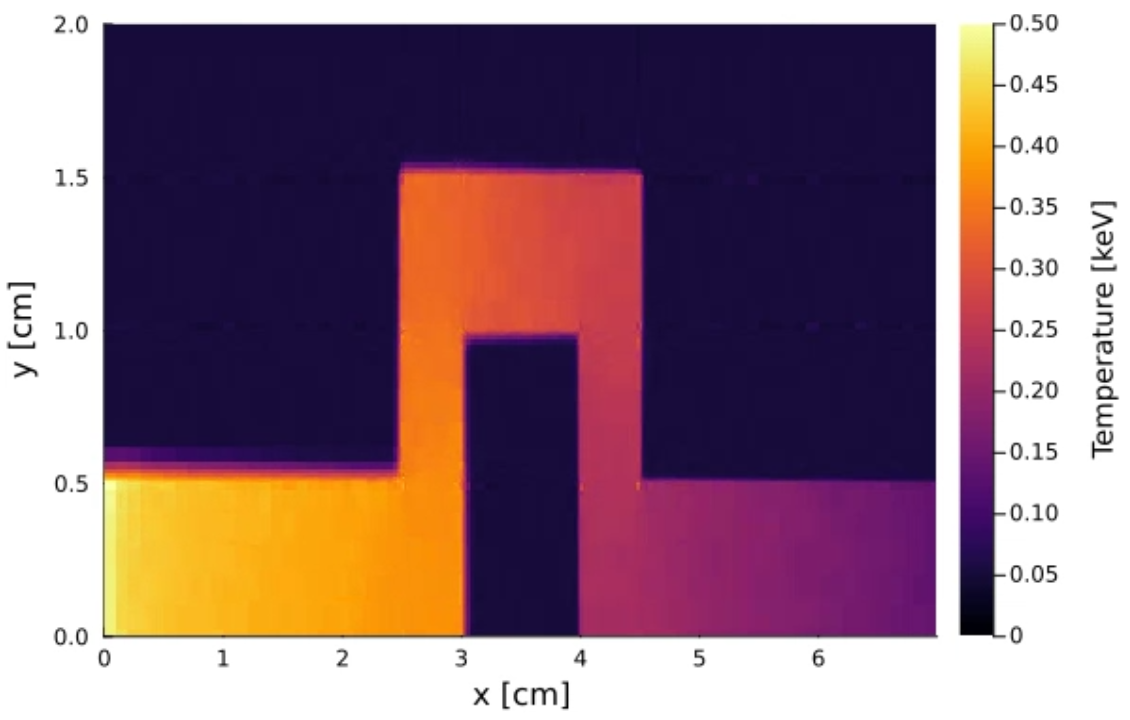}
    \caption{Crooked Pipe Half-Precision Heatmap at Time $t = 10.0$ Shakes}
    \label{fig:crookedpipe16}
\end{figure}

Both heatmaps show a similar distribution of temperature throughout the pipe at $t = 10.0$, with only minor single-cell temperature discrepancies being visible that are mainly attributable to the presence of statistical noise. The use of pairwise summation provides significant accuracy benefits to tallying energy deposition events in both optically thin and thick regions in half-precision. It is most effective in optically thick cells such as those at the pipe walls, where many small energy deposition events must be accurately summed to allow the cells to heat up and reradiate enough energy. Combined with the $\texttt{expm1}$ method, pairwise summation also prevents underheating from occurring in smaller area, optically thin cells where particles streaming through almost exclusively deposit very numerically small energies through implicit capture.

It is also standard in this benchmark problem to plot the temperature as a function of time at the fiducial points whose coordinates are given in Table \ref{tab:fiducialpoints} below:
\begin{table}[H]
\centering
\begin{tabular}{|c|c|c|}
\hline
\textbf{Fiducial Point} & \textbf{x (cm)} & \textbf{y (cm)} \\ \hline
\#1 & 0.0 & 0.25 \\ \hline
\#2 & 0.0 & 2.75 \\ \hline
\#3 & 1.25 & 3.5 \\ \hline
\#4 & 0.0 & 4.25 \\ \hline
\#5 & 0.0 & 6.75 \\ \hline
\end{tabular}
\caption{Crooked Pipe Fiducial Point Coordinates}
\label{tab:fiducialpoints}
\end{table}
The temperatures at these fiducial locations, ranging from time $t = 0$ to $t = 100$ shakes, were computed in double- and half-precision IMC simulations and are plotted in Figure \ref{fig:crookedpipefiducials}:

\begin{figure}[H]
    \centering
    \includegraphics[width=0.8\linewidth]{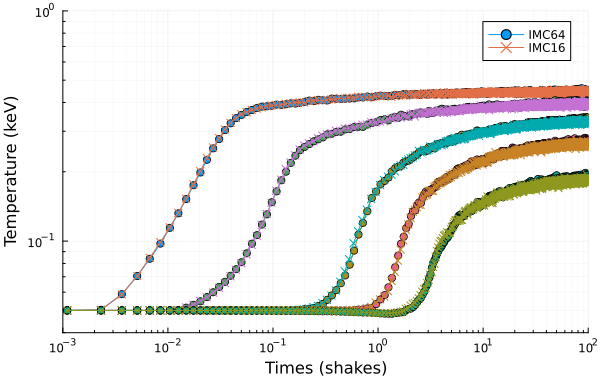}
    \caption{Crooked Pipe Double and Half-Precision Fiducial Point Temperatures vs Time}
    \label{fig:crookedpipefiducials}
\end{figure}
A curve for each fiducial point is included, with the leftmost data-series corresponding to point 1, the second leftmost data-series to point 2, and so on to the rightmost data-series for point 5. This plot shows that the temperatures at each point agree well between the double and half-precision cases. Overall, we see that with the judicious use of scaling and arithmetic manipulations, difficult 2D IMC problems like the Crooked Pipe can be modeled accurately in half-precision, albeit with a few minor issues. 

\section{Conclusions}
We have described and demonstrated the use of several techniques to enable accurate reduced-precision simulations of thermal radiation transport using the Implicit Monte Carlo Method. Arithmetic manipulations in the form of various numerical summation algorithms including naive summation, pairwise summation, and Kahan summation, as well as an algorithmic procedure to rearrange the order of operations for products are discussed. Scaling techniques, including both static rescaling of units as well as more dynamic scaling methods for quantities like distance with the aid of relative position tracking are shown. The idea of multiple-energy scaling was also introduced and demonstrated on the Crooked Pipe problem which could not be previously be modeled in half-precision. Results for a few different IMC benchmarks including the Su-Olson, the Crooked Pipe, and a Marshak Wave problem are shown for half and double precision implementations. The implementation of multiple energy scales and the more robust order of operations rearranging algorithm allows for accurate half-precision results to be generated for these problems that previous half-precision IMC efforts could not accurately solve. Additionally, the random walk diffusion acceleration approach was discussed for its ability to mitigate some of the issues presented by optically thick media in reduced-precision and shown to provide a performance increase in a selected problem. While this paper focuses on thermal radiation transport, the scaling and arithmetic tricks presented in this paper should be extensible to other types of scientific computation. Different types of reduced-precision calculations will all pose unique challenges, but prudent implementations of the reduced-precision techniques discussed in this paper will hopefully aid in the creation of accurate and performant reduced-precision codes. Future research efforts will involve improving the performance of reduced-precision computational techniques by implementing on actual reduced-precision hardware and extending their generality/usability for more widespread adoption.  

\section*{Acknowledgements}
This work was supported by the U.S. Department of Energy through the Los Alamos National Laboratory. Los Alamos National Laboratory is operated by Triad National Security, LLC, for the National Nuclear Security Administration of U.S. Department of Energy (Contract No. 89233218CNA000001).







\bibliography{doc/refs}
\bibliographystyle{elsarticle-num}
\end{document}